\documentstyle[12pt]{article}
\newcommand{\beq}{\begin{equation}}
\newcommand{\eeq}{\end{equation}}
\newcommand{\beqa}{\begin{eqnarray}}
\newcommand{\eeqa}{\end{eqnarray}}
\newcommand{\ba}{\begin{array}}
\newcommand{\ea}{\end{array}}
\newcommand{\CR}{\nonumber \\}
\newcommand{\pa}{\partial}
\newcommand{\A}{\alpha}
\newcommand{\B}{\beta}
\newcommand{\D}{\delta}          
\newcommand{\G}{\gamma}
\newcommand{\E}{\epsilon}

\newcommand{\lm}{\lambda}

\newcommand{\half}{{1\over 2}}

\newcommand{\Pf}{{\rm Pf}}

\newcommand{\cO}{{\cal O}}
\newcommand{\cL}{{\cal L}}
\newcommand{\hF}{{\hat F}}

\newcommand{\T}{\theta}

\begin{document}

\makeatletter
\def\setcaption#1{\def\@captype{#1}}
\makeatother

\begin{titlepage}
\null
\begin{flushright} 
hep-th/9911245  \\
UT-866  \\
November, 1999
\end{flushright}
\vspace{0.5cm} 
\begin{center}
{\Large \bf
Instantons in the $U(1)$ Born-Infeld Theory
and Noncommutative Gauge Theory
\par}
\lineskip .75em
\vskip2.5cm
\normalsize
{\large Seiji Terashima} 
\vskip 1.5em
{\large \it  Department of Physics, Faculty of Science, University of Tokyo\\
Tokyo 113-0033, Japan}
\vskip3cm
{\bf Abstract}
\end{center} \par
We derive a BPS-type bound for 
maximally rotational symmetric configurations in
four-dimensional Born-Infeld action 
with constant $B$ field background.
The supersymmetric configuration
saturates this bound and is regarded as
an analog of instanton in $U(1)$ gauge theory.
Furthermore, we find the explicit solutions of this BPS condition.
These solutions have a finite action 
proportional to the instanton number
and represent D$(p-4)$-branes within a Dp-brane
although they have a singularity at the origin.
Some relations to the noncommutative $U(1)$ instanton
are discussed.

\end{titlepage}

\baselineskip=0.7cm


In recent development of string theory,
it has been realized that 
noncommutative spacetime
is naturally appeared in the 
D-brane physics.
Indeed, it has been known that in a certain limit 
the transverse coordinates of D-branes can be regarded
as matrices.
This fact implies the non-commutativity of spacetime \cite{Wi1}.
Furthermore the noncommutative Yang-Mills theory is 
realized on the world volume of 
D-branes with constant $B$ field background.
This has been shown in the context of Matrix theory 
in \cite{CoDoSc} and more directly considering
open strings ending on the D-branes in \cite{DoHu,ChKr,KaOk}.

In \cite{SeWi}, Seiberg and Witten  
has promoted this idea deeper and argued that 
the noncommutative Yang-Mills theory is 
equivalent to the ordinary gauge theory.
They have also discussed the relations between
the ordinary instantons and 
the instantons in the noncommutative Yang-Mills theory 
\cite{NeSc} which has no small instanton singularities.
\footnote{It is discussed in \cite{Be} also that the relation between 
the instantons on branes 
and the noncommutative Yang-Mills theory.
The noncommutative instanton on the torus 
has been discussed in \cite{AsNeSc}.
The monopole in the noncommutative $U(2)$ Yang-Mills theory
has been considered in \cite{HaHaMo,HaHa} }
The instantons represent the D$(p-4)$-branes within the Dp-brane.

In the $U(1)$ case, 
the effective action on the brane for the slowly varying fields
has been known as the Dirac-Born-Infeld action 
\cite{Ts}. 
It has been shown in \cite{SeWi} that the 
BPS condition of the ordinary Dirac-Born-Infeld action and 
a noncommutative one
are equivalent in a limit $\A' \rightarrow 0$
with an appropriate rescaling of the metric.
The noncommutative $U(1)$ gauge theory has
instanton solutions \cite{NeSc} though 
the ordinary $U(1)$ gauge theory can not has
a nonsingular solutions with nonzero instanton numbers.
In the limit,
they have also constructed the BPS solutions with a finite instanton number
and a singularity at the origin.

In this paper we consider the BPS condition of
the ordinary Dirac-Born-Infeld action with $\A'$ fixed finite.
We derive the BPS-type bound for 
maximally rotational symmetric configurations in
the Born-Infeld action 
with constant $B$ field background
and show that the supersymmetric configuration indeed
saturates this bound.
Moreover we find the solutions of this BPS condition.
These solutions have a finite action 
proportional to the integral of $F \wedge F$
and may represent D$(p-4)$-branes in a Dp-brane
though they have a singularity at the origin.
Although these solution can not be valid because of the singularity,
we expect that the solutions is 
approximately valid away from the singularity.
Thus the $U(1)$ instanton solutions maybe
teach us about the relations between the two theories.


Let us consider the four-dimensional 
bosonic Born-Infeld Lagrangian, 
\beq
S=T \int{d x^4} \,  \sqrt{ \det \left( g+2 \pi \A' (F+B)  \right) },
\label{DBI}
\eeq
where $T$ is a constant \cite{Ts}
and parameterize the metric as 
\beq
g_{ij}= 2 \pi \A' \G \, \D_{ij}.
\eeq

First we discuss a BPS-type bound 
for the Born-Infeld action
with constant $B^+$ and $B^-=0$.
Here we concentrate on the gauge field configuration
with
\beq
F^+_{ij}=f(x) B^+_{ij},
\eeq
for simplicity.
This can be achieved by taking
the gauge potential to the form considered in \cite{SeWi},
\beq
A_i= B_{ij} x^j h(r),
\label{ai}
\eeq
and in \cite{SeWi} it has been stated that 
this is the most general 
$U(2)$-invariant ansatz 
where the $U(2)$ is a subgroup of $SO(4)$ that leaves $B^+$ fixed.

Defining 
$b^2=(B^+)^2/4$  and
\beqa
D_0&=& \G^2+b^2-\Pf F, \\
D_1&=& f+1  \\
D_2&=&  f+\frac{\Pf F}{\G^2+b^2}=D_1-\frac{1}{\G^2+b^2} D_0, 
\eeqa
we can show that
\beqa
\det \left[ \frac{1}{2 \pi \A'} g+ (F+B^+) \right] \!\!\!
&=&  \G^4+\half \G^2 (F+B)^2 +\left( \Pf (F+B) \right)^2  
\CR
&=& \! \G^4+2 \G^2  \left\{ b^2\! -\! \Pf F+2 f ( f  + 1) b^2 \right\}
\! +\!\left( b^2\!-\Pf F -2  b^2 (f+1) \right)^2 
\CR
&=& D_0^2+ 4 b^2 (\G^2+b^2) D_1 D_2 \CR
&=& (D_0+2 b^2 D_2)^2+4 b^2 \G^2 (D_2)^2 \geq (D_0+2 b^2 D_2)^2 ,
\eeqa
where we have used $F^2= F_{ij} F_{ij}=(F^+)^2+(F^-)^2$ and 
$4 \Pf F= F \tilde{F}=( (F^+)^2-(F^-)^2 )$.
Therefore using $S |_{F=0}=(2 \pi \A')^2 \, T (\G^2+b^2)$,
we obtain an inequality
\beq
S-S |_{F=0}\geq  (2 \pi \A')^2 \, T \int{d x^4} \, 
( -\Pf F +2 b^2 D_2).
\label{BPS}
\eeq
Here we simply have assumed that $D_0+2 b^2 D_2 \geq0$.
The inequality (\ref{BPS}) is a generalization of 
the BPS-type bound for DBI action with $B=0$ \cite{GaGoTo}
to the nonzero $B$ field.
This BPS-type bound is saturated by the configuration 
which satisfies the nonlinear equation $D_2=0$.
Later we will see that this equation is 
also a BPS condition to preserve supersymmetry 
if we consider the supersymmetric extension of $S$.
Note that $D_0+2 b^2 D_2$ is a constant plus a total derivative and 
depends only on the boundary value of the gauge field
because $f=F^+_{ij}/B^+_{ij}$.

Now we consider the four-dimensional Dirac-Born-Infeld action
which is an $N=4$ supersymmetric effective action for a D-brane 
if we ignore $\cO (\pa F)$ terms.
We can obtain $S$ by setting all scalars and fermions to zero.

Below we will restrict our attention to
the minimal $N=1$ supersymmetric extension of bosonic Born-Infeld 
action considered in \cite{BaGa},
which is a part of the Dirac-Born-Infeld action
since we expect that the other fields in full 
Dirac-Born-Infeld action
do not affect the analysis performed in this paper.

It is known that there are two distinct supertransformations for 
this action \cite{BaGa}.
One is a liner supertransformation given by
\beqa
\D \lm_{\A} &=& (F_{ij}^{+}+B_{ij}^{+}) \sigma^{ij \;\; \B }_{\;\; \A}
\eta_\B, \CR 
\D \bar{\lm}_{\dot{\A}} &=& (F_{ij}^{-}+B_{ij}^{-}) 
\sigma^{ij \;\; \dot{\B} }_{\;\; \dot{\A}}
\bar{\eta}_{\dot\B},
\eeqa
where $\eta$ is a parameter of the transformation.
The other one is nonlinear and broken if constant part of $F+B$ vanishes.
This is given by \cite{BaGa} \cite{SeWi},
\beqa
\D^* \lm_{\A} &=& \frac{1}{4 \pi \A'} 
\left( \sqrt{\det g} - (2 \pi \A')^2 \Pf(F+B)  
+\sqrt{ \det \left[ g+(2 \pi \A') (F+B) \right] } \right)
\eta_\A^* \CR 
&=& \pi \A' \left( \G^2 - \Pf(F+B)  
+\sqrt{ \G^4+\half \G^2 (F+B)^2 +\left( \Pf (F+B) \right)^2  } \right) 
\eta_\A^*, \CR 
\D^* \bar{\lm}_{\dot{\A}} &=& \frac{1}{4 \pi \A'} 
\left( \sqrt{\det g} + (2 \pi \A')^2 \Pf(F+B)  
+\sqrt{ \det \left[ g+(2 \pi \A') (F+B) \right] } \right)
\bar{\eta}_{\dot{\A}}^* \CR 
&=& \pi \A' \left( \G^2+ \Pf(F+B)  
+\sqrt{ \G^4+\half \G^2 (F+B)^2 +\left( \Pf (F+B) \right)^2  } \right) 
\bar{\eta}_{\dot{\A}}^*.
\eeqa

For $F=0$ and a constant $B$, it can be shown that
a combination of $\D+\D^*$ with
\beqa
\label{bps}
\eta_\A^* &=& -\frac{ B_{ij}^{+} }
{\pi \A' \left( \G^2- \Pf B  
+\sqrt{ \G^4+\half \G^2 B^2 +\left( \Pf B \right)^2  } \right) }
\, \sigma^{ij \;\; \B }_{\;\; \A} \, \eta_\B, \\
\bar{\eta}_{\dot{\A}}^* &=& -\frac{ B_{ij}^{-} }
{\pi \A' \left( \G^2+ \Pf B  
+\sqrt{ \G^4+\half \G^2 B^2 +\left( \Pf B \right)^2  } \right) }
\, \sigma^{ij \;\; \dot{\B} }_{\;\; \dot{\A}} \, \bar{\eta}_{\dot{\B}}, 
\eeqa
remains unbroken.

Below we will look for the BPS solution for 
the action
with a constant $B^+ \neq 0$ and $B^-=0$.
Then eq.(\ref{bps}) becomes 
\beqa
\eta_\A^* &=& -\frac{ B_{ij}^{+} }
{2 \pi \A'  \G^2}
\, \sigma^{ij \;\; \B }_{\;\; \A} \, \eta_\B, 
\label{bps1}
\eeqa
and the BPS condition $(\D+\D^* )\lm_\A=0$
is equivalent to
\beqa
\!\! 0 \!\! &=& \!\!\!\! F_{ij}^{+}+B_{ij}^{+} 
-\frac{1}{2 \G^2} B_{ij}^{+} 
\left( \G^2 - \Pf (F+B^+) \right. \CR
&& \left. \hspace{2.5cm}
+\sqrt{ \G^4+\frac{\G^2}{2} (F+B^+)^2
+\{ \Pf (F+B^+)\}^2  }
\right).
\label{bps2}
\eeqa
As a direct consequence of this, $F^+$ should be proportional to $B^+$
and we set 
\beq
F^+_{ij}=f(x) B^+_{ij}.
\label{fb}
\eeq
Then eq.(\ref{bps2}) becomes
\beq
2 \G^2 (f+1) -( \G^2-b^2-2fb^2-\Pf F )=
\sqrt{\{ \G^2-b^2-2fb^2-\Pf F \}^2+4 \G^2 (f+1)^2 b^2 }.
\label{bps3}
\eeq
Because the LHS of this equation is positive, 
we have solutions only if
\beq
2 \G^2 (f+1) -( \G^2-b^2-2fb^2-\Pf F )=
2 f (\G^2+b^2)+\G^2+b^2 +\Pf F  \geq 0,
\label{in}
\eeq
is satisfied. Then
the eq.(\ref{bps3}) is equivalent to 
\beqa
0 &=& 4 b^2 (f+1) 
\{ (f+1) (\G^2-b^2)+\left( -\G^2+b^2+2fb^2+\Pf F \right) \} \CR
&=& 4 b^2 (\G^2+b^2) D_1 D_2.
\eeqa
Therefore there are two types of  
solutions of the BPS condition.
The first one is $D_1=0$ with
$ D_0 \leq 0$.
The second one is 
\beq
\label{s2}
D_2=f+\frac{\Pf F}{\G^2+b^2}=0,
\eeq
with
\beq
D_0 \geq 0.
\label{s2a}
\eeq

To proceed further, we assume that the solution of (\ref{bps2}) has
the gauge field of the form (\ref{ai}).
We move to a frame where
\beq
B=\left(
\begin{array}{cccc}
0 & b & 0 & 0 \CR
-b & 0 & 0 & 0 \CR
0 & 0 & 0 & b \CR
0 & 0 & -b & 0
\end{array}
\right),
\label{B}
\eeq
without loss of generality.
We compute 
\beqa
\frac{F_{12}}{b} &=& -2 h-(x_1^2+x_2^2) \frac{h'}{r} \CR
\frac{F_{34}}{b} &=& -2 h-(x_3^2+x_4^2) \frac{h'}{r} \CR
\frac{F_{13}}{b} &=& \frac{F_{24}}{b}=(x_1 x_4-x_2 x_3) \frac{h'}{r} \CR
\frac{F_{23}}{b} &=& -\frac{F_{14}}{b}=(x_2 x_4-x_1 x_3) \frac{h'}{r},
\eeqa
which imply
\beqa
f &=&  -\frac{ 1}{2} \, ( 4 h + r h') ,  \\
\Pf F  &=& \left(\frac{b}{2}\right)^2  
\left( \left( 4 h + r h' \right)^2 
-( r h')^2 \right) = 2 b^2 h (2 h+ r h').
\label{pf}
\eeqa

Hence the solutions of the first condition
$0=D_1=-( 4 h + r h'-2)/2$
are $h=1/2+E / r^4$ where $E$ 
is some constant.
However, evaluating $D_0=\G^2+E^2/r^8 \geq 0$,
we find that this solution can not be accepted
unless $\G=E=0$.

To obtain solutions with $F$ decaying sufficiently
rapidly for large $r$,
we now solve the second condition (\ref{s2})
\beqa
0& =&f+\frac{\Pf F}{\G^2+b^2}=\frac{1}{\G^2+b^2}
\left( - \half (\G^2+b^2) ( 4 h + r h') + 2 b^2 h (2 h+ r h') \right) \CR
& =& 
\frac{2 b^2}{\G^2+b^2} 
\left[ \left( 2 h^2 -\left(1+\frac{\G^2}{b^2} \right) h 
\right)
+\frac{1}{4} r \frac{d}{dr} \left(
2 h^2 -\left(1+\frac{\G^2}{b^2} \right) h 
\right) \right].
\eeqa
The solution of this differential equation is given by
\beq
2 h^2 -\left(1+\frac{ \G^2}{b^2} \right) h = \frac{4 N}{b^2 \, r^4},
\eeq
where $N $ is some dimensionless constant.
Therefore the BPS solution is given by
(\ref{ai}) with 
\beqa
h & =& \frac{1}{4} \left( 
\left(1+\frac{ \G^2}{b^2} \right)
-\sqrt{\left(1+\frac{ \G^2}{b^2} \right)^2+\frac{32N }{b^2 \, r^4}  }
\right) \CR
& =& \frac{1}{4} \left(1+\frac{ \G^2}{b^2} \right)
\left( 1-\sqrt{1+\frac{r_c^4}{r^4} } \right),
\label{sol}
\eeqa
where 
\beq
r_c= \left( 32 N \frac{b^2 }{(\G^2+b^2)^2}  \right)^{\frac{1}{4}}.
\eeq
Note that  the gauge potential and the field strength have
a singularity at $r=0$ and
$N$ does not have to be quantized in this analysis.
Later we will shortly discuss these problems.
We also note 
that if this solution satisfies the condition (\ref{s2a}), 
it is also a solution of the equation of motion of
bosonic Born-Infeld action
with a constant $F^+$ background except at the origin.

To show that this solution satisfies (\ref{s2a}),
remember the formula
\beq
\Pf F=b^2 \frac{1}{r^3} \frac{d}{dr} (r^4 h^2).
\eeq
Using this formula,
we can see that $\Pf F  \rightarrow 0$ 
in a limit $r \rightarrow \infty$
and
\beqa
\frac{d}{dr} D_0 
&=&-b^2 \frac{d}{dr} \frac{1}{r^3} \frac{d}{dr} (r^4 h^2)
=-\half \frac{d}{dr} \frac{1}{r^3} \frac{d}{dr}
\left[ r^4 \left\{ (\G^2+b^2) h+\frac{4N}{r^4} \right\} \right] \CR
&=& - \half  (\G^2+b^2) \frac{d}{dr} \left( 4 h +r h' \right)
=- \half  \frac{
(\G^2+b^2) \left( \frac{32 N}{b^2 r^5} \right)^2 }{
\left(  \left(1+\frac{ \G^2}{b^2} \right)^2
+\frac{32N }{b^2 \, r^4}
\right)^{\frac{3}{2} }  } \CR
& \leq & 0.
\eeqa
From this, we conclude $D_0  > 0$
for the solution satisfying (\ref{sol}).
According to (\ref{sol}), however,
the gauge potential has a nonzero imaginary part 
if $N<0$ and $ \left(1+\frac{ \G^2}{b^2} \right)^2
<\frac{-32N }{b^2 \, r^4}$.
Thus we conclude that 
the solution (\ref{ai}) with (\ref{sol}) is consistent
if and only if $N>0$.

We can see that the instanton number is $N$ from
the computation
\beqa
\int d^4 x F \tilde{F} &=&  2 \pi^2 \int_0^{\infty}
r^3 dr \, b^2 \, 8 h \, (2 h+r h')=8 \pi^2 b^2
\int_0^{\infty} dr \, \frac{d}{dr} (r^4 h^2) \CR
&=& - 8 \pi^2 b^2 \lim_{r \rightarrow 0} (r^4 h^2) \CR
&=& -16 \pi^2  N.
\eeqa
Although the solution has the singularity at $r=0$, 
its action is finite and proportional to the instanton number.
This follows from the BPS-type bound (\ref{BPS}) and $D_2=0$.
Indeed we obtain
\beqa
S-S |_{F=0} &=& T \int{d x^4} \,  \left\{ 
\sqrt{ \det \left( g+2 \pi \A' (F+B)  \right) }-
\sqrt{ \det \left( g+2 \pi \A' (B)  \right)  }
\right\} \CR
&=&  T (2 \pi \A')^2 4 \pi^2 N.
\eeqa 
If we consider a D$p$-brane, $T$ becomes its tension
and the instantons represents D$(p-4$)-branes within it
\cite{Do,Wi}.
In this case, the effective Lagrangian contains
a Chern-Simons term
\beq
\cL_{CS}=\frac{1}{2 (2 \pi)^2 } \int_{Dp} C_{p-3} \wedge F \wedge F,
\eeq
where $C_{p-3}$ is a RR $(p-3)$-form.
In the instanton background,
this induces a coupling 
\beq
\cL_{CS}=-N \int_{D(p-4)} C_{p-3},
\eeq
which implies that $N$ is the number of the D$(p-4$)-branes.
Since the tension of a D$p$-brane is 
$T_p=1/(g_s (2 \pi )^p (\A')^{\frac{p+1}{2}} ) $,
the action of the instantons within the brane is given by
\beq
S-S |_{F=0}=
\frac{N}{ g_s (2 \pi )^{p-4} (\A')^{\frac{p-3}{2}} },
\eeq
and agrees with the tension of the D$(p-4)$-brane.

Here we should require $N$ to be an integer number.
It is possible that this quantization of $N$ 
is required
if we include the $\cO (\pa F)$ terms into the action.
However even if this is indeed so and considering the $r \gg r_c$ region,
$N$ defined in (\ref{sol}) can be corrected
to become a non-integer number
by inclusion of these terms.

Now we study the solutions in various limits.
First in the $r \gg r_c $ region,
the solution becomes
\beq
h \sim -\frac{\G^2+b^2}{8 b^2} \left( \frac{r_c}{r} \right)^4,
\eeq
and in the $r \ll r_c $ region it becomes
\beq
h \sim -\frac{\G^2+b^2}{4 b^2} \left( \frac{r_c}{r} \right)^2.
\eeq

The zero slope limit of \cite{SeWi},
$\A'\sim \E^{\half}, g_{ij} = \E \D_{ij}$ with $\E \rightarrow 0$,
corresponds to $\G \rightarrow 0$.
In this limit the corresponding action of the noncommutative theory
becomes the $\hF^2$ action and the BPS condition becomes 
$F^+=0$.
This condition has been matched with eq. (\ref{bps2})
in the limit.
The solutions (\ref{sol}) 
agree with the solutions obtained in \cite{SeWi}
in the $\G \rightarrow 0$ limit.

If we take $B \rightarrow 0$ limit,
the instantons are expected to be very small to be able to
escape from the Dp-brane,
as the small instanton singularity in the 
world volume non-Abelian gauge theory of Dp-branes.
Indeed, we can easily show that in the $b \rightarrow 0$ limit,
$r_c^2 \sim b/ \G^2$ and 
$F \rightarrow  \frac{b}{\G^2} \frac{1}{r^4}$ 
in the region $r \gg r_c $.
Thus in this limit 
the instantons are localized 
in the very small region with the length scale $r_c
(\frac{\G}{b})^{s}$
with $1/4 <s<1/2$ and the singularity becomes worse.
The usual $\A' \rightarrow 0$ limit corresponds to 
$\G \rightarrow \infty$. In this limit also, the instantons 
become small instantons.

We expect that in the region $r \gg r_c$
the solutions obtained in this paper is
valid because $\pa F$ is very small.
However, 
since the solutions vary too rapidly near the origin,
the effects of the $\cO (\pa F)$ terms can not 
be neglected.
The solutions may become nonsingular
due to the correction from these terms
and correspond to the nonsingular noncommutative $U(1)$ 
instantons.

Finally we represent the natural parameters $G$ and $\T$ in 
the noncommutative gauge theory by the ones in ordinary gauge theory.
Denoting $G_{ij}= 2 \pi \A' \tilde{\G} \D_{ij}$ and
\beq
\T=-(2 \pi \A')^2 (g+2 \pi \A' B)^{-1} B (g-2 \pi \A' B)^{-1}  
=\left(
\begin{array}{cccc}
0 & t & 0 & 0 \CR
-t & 0 & 0 & 0 \CR
0 & 0 & 0 & t \CR
0 & 0 & -t & 0
\end{array}
\right),
\eeq
we find
\beqa
\tilde{\G}&=&\frac{G}{2 \pi \A'}=\frac{g}{2 \pi \A'}-(2 \pi \A') (B g^{-1} B)
=\frac{\G^2+b^2}{\G} , \\
t &=& - \frac{b}{b^2+\G^2} =- \frac{1}{\sqrt{32 N}} r_c^2.
\eeqa
We note that $r_c^2$
is
roughly identified with the scale of noncommutativity $t$.


In conclusion we have derived the BPS-type bound for 
the maximally rotational symmetric configurations in
four-dimensional Born-Infeld action 
with constant $B$ field background.
The supersymmetric $U(1)$ instanton configuration
saturates this bound.
We have found the explicit solutions of this BPS condition.
Although this solution has a singularity at the origin,
but this has 
a finite action which is consistent with
the interpretation of this as D(p-4) branes within Dp-branes.
Some relations to the noncommutative $U(1)$ instanton
has been discussed.
To further explore the connection between two theories
it will be interesting to
investigate the BPS solution of the 
world volume theory of the multiple Dp-branes with constant $B$ field
because in this case the usual non-Abelian instanton
is not singular.

\vskip6mm\noindent
{\bf Acknowledgements}

\vskip2mm
I am grateful to the organizers of the workshop
Summer Institute '99 at Fuji-Yoshida.
I would like to thank J. Hashiba, K. Hashimoto, K. Hosomichi, T. Kawano,
Y. Okawa and K. Okuyama for useful discussions.
This work was supported in part by JSPS Research Fellowships for Young 
Scientists. \\

\noindent
{\bf Note added}: 

While preparing this article for publication, we received the preprint
\cite{MaMiMoSt} which gives derivations of 
the BPS condition 
\beq
\frac{(F+B)^+}{\Pf (F+B)-\G^2}=\frac{B^+}{\Pf B-\G^2},
\eeq
in our notation, which is equivalent to (\ref{fb}) and (\ref{s2}).

\newpage


\end{document}